\documentclass{aa}

\usepackage{graphicx}
\usepackage{aabib99}

\newcommand{\La}{\mbox{${\rm Ly\alpha}$}}
\newcommand{\Nh}{\mbox{$N_{\rm HI}$}}
\newcommand{\Line}[3]{\Ion{#1}{#2}\,$\lambda$\,#3}

\newcommand{\Ion}[2]{#1{\,\scriptsize #2}}
\newcommand{\Teff}{\mbox{$T_{\rm eff}$}}
\newcommand{\Msun}{\mbox{$M_{\odot}$}}
\newcommand{\es}{\mbox{$\rm erg\;s^{-1}$}}

\begin{document}

\thesaurus{06   
          (02.01.2;         
           08.09.2 RX\,J0439.8-6809; 
           08.14.2;         
           08.23.1;         
           13.25.5)}        

\title{HST/STIS ultraviolet spectroscopy of the 
       supersoft X-ray source RX\,J0439.8$-$6809
\thanks{Based on observations made with the NASA/ESA Hubble Space
Telescope, obtained at the Space Telescope Science Institute, which is
operated by the Association of Universities for Research in Astronomy,
Inc., under NASA contract NAS 5-26555.}}

\author{A. van Teeseling\inst{1} \and
        B.T. G\"ansicke\inst{1} \and
        K. Beuermann\inst{1} \and
        S. Dreizler\inst{2} \and
        T. Rauch\inst{2} \and 
        K. Reinsch\inst{1}}

\offprints{B. G\"ansicke, boris@uni-sw.gwdg.de}

   \institute{  Universit\"ats-Sternwarte G\"ottingen,
                Geismarlandstr. 11, D-37083 G\"ottingen, Germany
        \and    Institut f\"ur Astronomie und Astrophysik,
                Waldh\"auser Str. 64, D-72076 T\"ubingen, Germany
             }
\date{Received 26 August 1999 / Accepted 11 October 1999}

\maketitle

\begin{abstract}
We present ultraviolet observations of the supersoft X-ray source
RX\,J0439.8$-$6809 obtained with the Hubble Space Telescope Imaging
Spectrograph.
The ultraviolet spectrum is a very blue continuum
overlayed with interstellar absorption lines.  The observed broad \La\
absorption line is consistent with an interstellar column density of
neutral hydrogen $\Nh=(4.0\pm1.0)\times10^{20}\rm\;cm^{-2}$.
The light curve obtained from the time-tagged dataset puts a $3\sigma$
upper limit of 0.04\,mag on the ultraviolet variability of
RX\,J0439.8$-$6809 on time scales between 10\,s and 35\,min.
The long-term X-ray light curve obtained from our three-year
ROSAT\,HRI monitoring of RX\,J0439.8$-$6809 shows the source with a
constant count rate, and implies that the temperature did not change
more than a few 1000\,K. 
If RX\,J0439.8$-$6809 is a massive extremely hot pre-white dwarf on
the horizontal shell-burning track, opposed to the alternative
possibility of a very compact double-degenerate supersoft X-ray
binary, its constant temperature and luminosity are a challenge to
stellar evolution theory. Interestingly, RX\,J0439$-$6809 is found
close to the theoretical carbon-burning main-sequence.

\end{abstract}

\section{Introduction}
Among the optically identified persistent supersoft X-ray sources,
RX\,J0439.8$-$6809 (hereafter RX\,J0439) has the largest X-ray
luminosity, but  is also  the faintest one in the visual
light. After its discovery with ROSAT \cite{greineretal94-1} and its
optical identification with a faint blue star in the LMC
\cite{vanteeselingetal96-1}, we are still puzzled by its exotic
nature. The ROSAT X-ray spectrum of RX\,J0439 is consistent with an
object in the LMC with a radius of a few times $10^9$\,cm, an
effective temperature of $3\times 10^5$\,K, and a luminosity of $\sim
10^{38}$\,erg\,s$^{-1}$. These parameters suggest that RX\,J0439 is a
shell-burning white dwarf or pre-white dwarf.
Two scenarios have been suggested by van Teeseling et
al. \cite*{vanteeselingetal97-1} that could explain the observed
characteristics of RX\,J0439.

(a) RX\,J0439 is a supersoft X-ray binary, and the high X-ray
luminosity is powered by stable nuclear shell burning. The absence of
any X-ray or optical variability in RX\,J0439, combined with its
optical faintness, excludes the presence of a quasi-main-sequence
companion. If RX\,J0439 belongs to the supersoft X-ray binaries, it
must be a double-degenerate system containing two semi-detached white
dwarfs, with an orbital period of a few minutes only. In this case,
binarity is extremely hard to prove: with no or only a tiny accretion
disc and a faint degenerate companion, the flux is dominated by the
accreting shell-burning white dwarf. This model could explain the
absence of strong emission lines which have been observed in all other
known supersoft X-ray binaries (see e.g. overview by van Teeseling
1998\nocite{vanteeseling98-1}), as well as the fact that the optical
spectrum of RX\,J0439 is consistent with the Rayleigh-Jeans tail of
the X-ray component.

(b) RX\,J0439 is an exceptionally hot pre-white dwarf on the
horizontal shell-burning track. In this case, RX\,J0439 would be the
hottest star of this type known so far.
However, to match the observed X-ray luminosity, such a pre-white
dwarf would have to be rather massive, with accordingly short
evolution times.

\section{HST/STIS observations}
HST/STIS ultraviolet observations of RX\,J0439 were carried out on
1998 November 17.  Due to the faintness of the object, the target
acquisition was performed on a nearby bright star, and RX\,J0439 was
positioned in the $52\arcsec\times0.5\arcsec$ slit with a subsequent
offset.
A 2100\,s exposure far-UV  spectrum (FUV, 1150$-$1730\,\AA) and a
1700\,s near-UV spectrum (NUV, 1600$-$3200\,\AA) were obtained with
the G140L and G230L gratings, respectively.
Using the MAMA detectors in the time-tagged mode, we have obtained in
addition to the ultraviolet spectra two photon event files which contain an
entry ($t, x, y$) for each photon, with $t$ the arrival time at a
$125\mu\,\rm s$ time resolution and $x, y$ the detector coordinates.
In order to extract light curves of RX\,J0439 from the two
observations we proceeded as follows.  Two-dimensional raw FUV
and NUV detector images were obtained by summing up all photons
registered in each individual pixel ($x, y$).
FUV source+background counts were extracted within a box 35\,pixel
wide in cross-dispersion direction and covering
$1230\,\mathrm{\AA}\la\lambda\la1720\,\mathrm{\AA}$, excluding the
strong geocoronal \La\ emission (but including the weak
\Line{O}{I}{1302} airglow).
Background counts were extracted from two adjacent empty regions on
the detector with the same wavelength coverage as the source
spectrum. From the NUV photon event file, we extracted
source+background and background counts covering the entire observed
wavelength range in a similar way.

\begin{figure}
\includegraphics[angle=270,width=8.8cm]{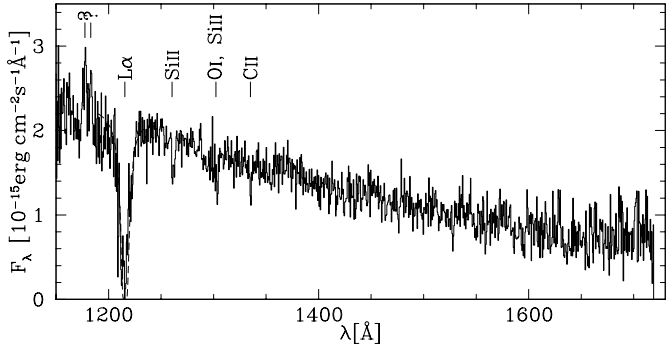}
\vspace*{-2ex}
\caption[]{The interstellar \La\ absorption line in the G140L spectrum
of RXJ\,0439. Plotted as a dashed line is a  Lorentzian 
fit to the observed line with a column density of
$\Nh=4\times10^{20}\rm\;cm^{-2}$. The absorption lines of \Ion{Si}{II},
\Ion{C}{II}, and \Ion{O}{I} are likely of interstellar origin. Two
unidentified possible emission features at $\approx1178$\,\AA\ and
$\approx1184$\,\AA\ are indicated}
\label{f-g140lspectrum}
\vspace*{-1ex}
\end{figure}

\section{The ultraviolet spectrum}
The ultraviolet spectrum of RX\,J0439 is a blue continuum which
contains a broad \La\ absorption line and weak absorption features at
$\sim1260$\,\AA, $\sim1300$\,\AA, and $\sim1335$\,\AA, which we
identify as interstellar absorption of \Line{Si}{II}{1260},
\Line{O}{I}{1302}/\Line{Si}{II}{1304}, and \Line{C}{II}{1335}
(Fig.\,\ref{f-g140lspectrum}).  Some spurious structure remains in the
\La\ profile and around 1300\,\AA\ after the subtraction of the
airglow \La\ and \Ion{O}{I} emission.  The observed \La\ absorption is
centred at $\approx1217$\,\AA.  Due to the offset target acquisition,
RX\,J0439 might not have been perfectly centred in the aperture, which
could account for the error in the wavelength zero point.  The
equivalent widths measured from the three metal absorption features,
$750\pm300$\,m\AA, $450\pm200$\,m\AA, and $500\pm200$\,m\AA,
respectively, are compatible with just the galactic foreground
absorption (e.g. \nocite{gaensickeetal98-1}G\"ansicke et al. 1998) and
make it plausible that RX\,J0439 is located relatively far in
the outskirts of the LMC \cite{vanteeselingetal96-1}.

Interpreting the observed \La\ profile as interstellar absorption, we
have estimated the neutral hydrogen column density along the line of
sight towards RX\,J0439.  Using a pure damping profile we find $\Nh =
(4\pm 1)\times 10^{20}$\,cm$^{-2}$. This value is lower than those
found for CAL\,83 and RX\,J0513.9$-$6951 \cite{gaensickeetal98-1} and is
consistent with the estimated galactic foreground column density of
$\Nh=4.5\times10^{20}$\,cm$^{-2}$ \cite{dickey+lockman90-1}. As for
the interstellar metal lines, there can be no significant \La\
absorption by interstellar or circumstellar material in the LMC. The
upper limit on the reddening derived from the G230L spectrum,
$E_\mathrm{B-V}\la0.1$, is consistent with the \Nh\ column density.
The derived value for \Nh\ is also consistent with the relatively low
absorption column found from the ROSAT X-ray spectrum
\cite{vanteeselingetal96-1}, and we will use $N_{\rm H} = 4\times
10^{20}$\,cm$^{-2}$ for the total hydrogen column towards RX\,J0439
throughout the rest of this paper.

The absence of \Line{N}{V}{1240} and \Line{He}{II}{1640} emission,
which has been observed in CAL\,83, CAL\,87, and RXJ\,0513.9$-$6951
\cite{gaensickeetal98-1,hutchingsetal95-1}, reminds of  the
pure continuum spectra observed in the optical.
However, the rather noisy blue end of the G140L spectrum of RX\,J0439
contains two possible emission features, which are centred at
$\approx1178$\,\AA\ and $\approx1184$\,\AA\ and are detected at a
$2-3\sigma$ level.
Possible identifications are \Line{C}{III}{1175.7} and
\Line{C}{IV}{1184.7}, even though two reasons argue against these
transitions. (1) If \Ion{C}{IV} were present in the atmosphere of
RX\,J0439, the resonance line \Line{C}{IV}{1550} should be much
stronger than \Line{C}{IV}{1184.7}. (2) \Ion{C}{III} is increasingly
less populated for temperatures in excess of 120\,000\,K, which is
much lower than the temperature derived below from the overall
spectrum. Such a low temperature, however, could be found on the
irradiation-heated surface of an accretion disc or degenerate
secondary star in the case of a (so far unproven) binary nature of
RX\,J0439.

\begin{figure}
\includegraphics[angle=270,width=8.8cm]{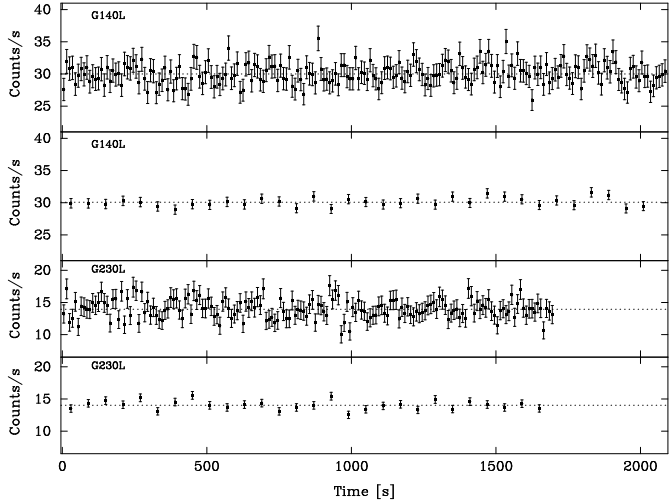}
\vspace*{-1.5ex}
\caption[]{Count rate of RX\,J0439 during the STIS G140L and G230L
observations in 10\,s and 60\,s bins}
\label{f-uvvar}
\vspace*{-1ex}
\end{figure}

\section{The absence of ultraviolet variability}
Figure~\ref{f-uvvar} shows the background subtracted count rate of
RX\,J0439 binned in 10\,s and 60\,s intervals for both
gratings. Neither observation shows significant variability. To
determine the upper limit to any random variability we have performed
a Monte Carlo simulation using the observed errors: The G140L
observation gives with a $3 \sigma$ upper limit of
$1.0$\,count\,s$^{-1}$ (corresponding to 0.04\,mag) the strongest
constraint on possible ultraviolet variability. The G230L observation
gives a $3 \sigma$ upper limit of $1.7$\,count\,s$^{-1}$
(corresponding to 0.13\,mag).

\begin{figure}
\includegraphics[angle=270,width=8.8cm]{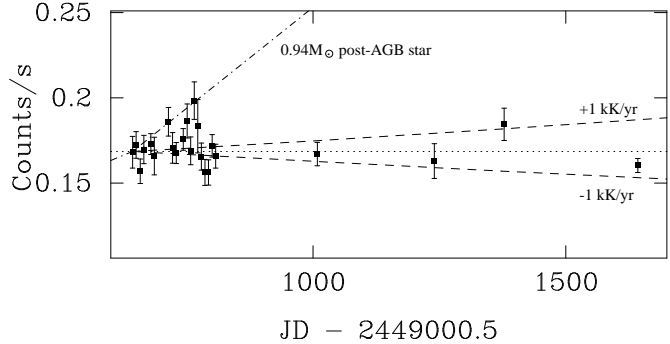}
\vspace*{-1.5ex}
\caption[]{Long-term X-ray light curve of RX\,J0439 obtained with the
ROSAT HRI.  The dashed lines show the expected change in count rate
with a secular change of $\pm 1\,000$\,K\,yr$^{-1}$ at constant
bolometric luminosity and with an effective temperature close to
270\,000\,K. The steeply rising dash-dotted line shows the
evolutionary change of a $0.94 M_{\sun}$ pre-white dwarf with roughly
the correct temperature and luminosity \cite{bloecker95-1},
corresponding to a change of $\sim 10\,000$\,K\,yr$^{-1}$ at nearly
constant bolometric luminosity}
\label{f-hri}
\vspace*{-1ex}
\end{figure}

The upper limit of $0.04$\,mag in the ultraviolet is even more
stringent than the  $3 \sigma$ upper limit of $0.07$\,mag in the
optical (van Teeseling et al. 1997). Combined with a visual magnitude
of $V=21.6$, the absence of any optical or ultraviolet variability
excludes an irradiated companion with the size of a main-sequence
star. Although the flux in a double-degenerate supersoft X-ray binary
is dominated by the expanded shell-burning primary, a small
quasi-sinusoidal modulation is expected from the changing aspect of
the irradiated degenerate helium donor star.
The stringent upper limit on the ultraviolet variability, however,
implies a low orbital inclination (even in the case of a
double-degenerate supersoft X-ray binary), or a rather ineffective
heating of the degenerate companion, or no companion at all, i.e. a
single pre-white dwarf.
\footnote{A substantial UV modulation as the result of reprocessing on
the secondary {\it is} expected and observed in double-degenerates
containing a {\it neutron star} and a brown dwarf (e.g. Arons \& King
(1993); Anderson et al. 1997\nocite{arons+king93-1,andersonetal97-1}).
In these LMXBs, however, the neutron star contributes only little to
the observed UV flux, while in RX\,J0439 the shell-burning white dwarf
is the dominant source of UV radiation.}

\section{The absence of long-term X-ray variability}
In Fig.~\ref{f-hri} we have plotted the count rate of RX\,J0439 from
our ROSAT\,HRI monitoring of RX\,J0439. The 25 pointings
cover a period of almost 3 years.  The count rate is not significantly
variable, and the data exclude a temperature change (at constant
bolometric luminosity) larger than $\pm 1\,000$\,K\,yr$^{-1}$.

\begin{figure}
\includegraphics[angle=270,width=8.8cm]{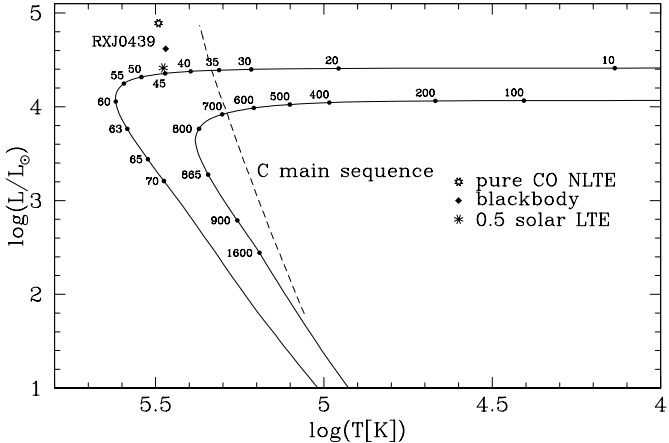}
\vspace*{-2ex}
\caption[]{RX\,J0439 in the Hertzsprung-Russell diagram. Indicated are
the results for ($L,T$) from three different fits to the overall
spectrum, using a pure CO NLTE model spectrum, a blackbody, and a 0.5
solar abundance LTE model spectrum. Plotted as solid lines are
evolutionary tracks for 0.696\,\Msun\ ($M_{\rm ZAMS}=4\,\Msun$) and
0.940\,\Msun\ ($M_{\rm ZAMS}=7\,\Msun$) white dwarfs from
\cite{bloecker95-1}. Evolutionary times are given in years. The dashed
line is the theoretical carbon main sequence
\cite{kippenhahn+weigert90-1}.}
\label{f-hrd}
\vspace*{-1ex}
\end{figure}

Comparison of the temperature and luminosity of RX\,J0439 (Sect.\,5)
with calculations of evolutionary tracks of post-AGB stars
(e.g. Bl\"ocker 1995\nocite{bloecker95-1}) shows that if RX\,J0439 is
a pre-white dwarf on the horizontal shell-burning track, its mass must
be $\ga0.9$\Msun\ (Fig.\,\ref{f-hrd}).  However, these massive white
dwarfs evolve so fast near the turn-over to the white dwarf cooling
track that even in the short history of ROSAT observations of
RX\,J0439 we should have seen a significant increase in the HRI count
rate (note that on the horizontal shell-burning track the effective
temperature increases at nearly constant bolometric luminosity). A
lower mass white dwarf [e.g.  the ($M_{\rm ZAMS}$, $M_{\rm H}$) =
(5\,\Msun, 0.836\,\Msun) track of \nocite{bloecker95-1}Bl\"ocker 1995]
evolves slow enough to be consistent with the available X-ray data,
but has a luminosity of $\la 5\times 10^{37}$\,\es\ at the appropriate
temperatures, which is significantly lower than that of RX\,J0439.
This implies either that the shell-burning in RX\,J0439 is powered by
accretion or that RX\,J0439 is able to stay at its position in the
Hertzsprung-Russell diagram by nuclear burning with a much longer
lifetime than predicted by the evolutionary tracks. With respect to the
last possibility it should be noted that RX\,J0439 is rather close to
the theoretical carbon-burning main sequence.

\section{The combined X-ray, ultraviolet and optical spectrum}
Our ultraviolet spectrum of RX\,J0439 perfectly agrees both in flux
and in slope with the observed optical spectrum presented by van
Teeseling et al. \cite*{vanteeselingetal96-1}. The combined spectrum
has a very blue slope consistent with the Rayleigh-Jeans tail of a
very hot object, and it is possible to model the entire observed
spectrum from X-rays to optical with a single optically thick
component.  This does not imply, however, that we can exclude
additional flux in the ultraviolet and optical, provided that the
additional flux has a very blue spectrum as well. Such additional flux
is required if the supersoft X-ray component has a higher temperature
(and consequentially a smaller inferred radius and bolometric
luminosity) than derived with the assumption that the ultraviolet flux
is the Rayleigh-Jeans tail of this X-ray component.

\begin{figure}
\includegraphics[width=8.8cm]{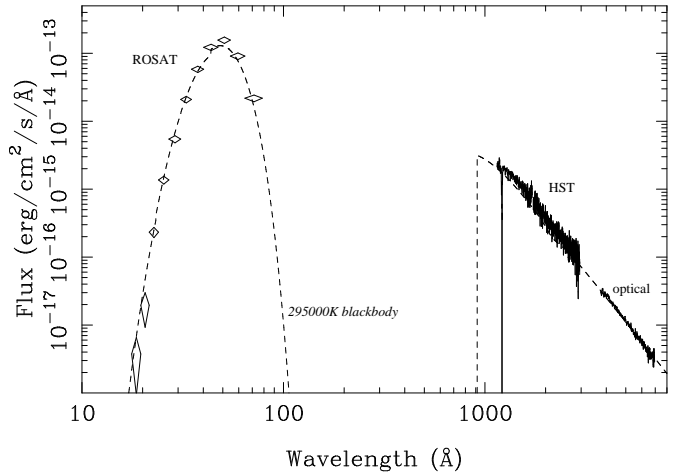}
\vspace*{-2ex}
\caption[]{Observed optical, ultraviolet and {\em deconvolved} X-ray
spectrum of RX\,J0439. The dashed line is an absorbed 295\,000\,K
blackbody spectrum}
\label{f-xuvopt}
\vspace*{-1ex}
\end{figure}

It is striking how well the overall spectrum of RX\,J0439, including
the absence of any detectable spectral features, matches a single
absorbed blackbody with a temperature of $295\,000$\,K
(Fig.~\ref{f-xuvopt}). If we assume a distance of 50\,kpc, this
blackbody gives a radius $R = 5\times 10^9$\,cm and a bolometric
luminosity $L = 1.6\times 10^{38}$\,\es. Because the lack of spectral
features, in particular in the ROSAT X-ray spectrum, may be the result
of the rather limited spectral resolution and signal-to-noise ratio,
we follow van Teeseling et al. \cite*{vanteeselingetal96-1} and
have fitted $\log g = 7$ white dwarf model atmospheres to the combined
X-ray, ultraviolet and optical spectrum (in fact a $\chi^2$ fit to the
ROSAT spectrum with the demand that the Rayleigh-Jeans tail matches
the observed ultraviolet and optical flux). In addition to LTE spectra
\cite{vanteeselingetal94-1}, we have now also used NLTE white dwarf
spectra \cite{rauch97-1} in order to investigate which ultraviolet
lines  might be expected to be present.

Van Teeseling et al. \cite*{vanteeselingetal96-1} already argued that
the atmosphere must contain a significant amount of metals. If not,
there would be either an excess of soft X-ray flux or an excess of
ultraviolet and optical flux.  Even models with 10\% solar metal
abundance suffer this problem.  With metal abundances within a factor
of 2 of solar, both the LTE and the NLTE models give an acceptable fit
to the overall spectrum with $\Teff\approx 300\,000$\,K and $L\approx
1\times 10^{38}$\,\es.  Because all models are scaled to the same
optical flux, the resulting fit parameters of the LTE and NLTE models
do not differ significantly. The near-solar models predict strong
\Ion{Ne}{VIII} absorption edges at 0.22\,keV and 0.24\,keV.  
For a cosmic helium abundance, the ultraviolet spectra predict a
non-negligible \Line{He}{II}{1640} absorption line which is,
however, not detected at the present signal-to-noise ratio.
This implies either that the assumption of a single spectral component
is incorrect, or that the absorption is filled in with emission
(without extra continuum flux), or that the white dwarf atmosphere is
helium-poor. Note that also in the optical no \Line{He}{II}{4686}
could be detected and that it appears unlikely that absorption is
filled in without the appearance of emission lines

Metal rich and helium poor is reminiscent of the exotic and extremely
hot ($\sim200\,000$\,K) PG\,1159 star H1504+65, which is the only
known pre-white dwarf whose surface is free from both hydrogen and
helium \cite{werner91-1,werner+wolff99-1}.  Indeed, the overall
spectrum of RX\,J0439 can be fitted very well (i.e. best $\chi^2$ of
all used models) with a single $\log g = 7$ spectrum with a pure CO
composition. The inferred effective temperature is $\sim 310\,000$\,K,
the radius at 50\,kpc is $\sim 6\times 10^9$\,cm, and the
corresponding luminosity is $\sim 3\times10^{38}$\,\es.
At this point, we can only speculate about further common properties
of the two stars.  In contrast to the featureless optical continuum of
RX\,J0439, the optical spectrum of H1504+65 contains numerous high
excitation C and O lines. VLT and Chandra observations of RX\,J0439
are scheduled to probe for emission/absorption features which will allow
a more detailed spectral modelling.

\section{Conclusions}
Our single orbit of HST/STIS ultraviolet data of RX\,J0439 have
confirmed the previous findings: RX\,J0439 is a very extreme object,
whether as a single post-AGB star or as an accreting
(double-degenerate) supersoft X-ray binary.  It seems that the flux at
all wavelengths is dominated by a very luminous shell-burning white
dwarf which is able to maintain its position in the
Hertzsprung-Russell diagram near the turnover into the white dwarf
cooling track for $>8$ years. The combined X-ray, ultraviolet and
optical spectrum is consistent with a single spectral component with
$T\sim 300\,000$\,K and $L\sim 10^{38}$\,erg\,s$^{-1}$, and suggests a
high metalicity. Interestingly, both the very good fit with a pure CO
model, the absence of long-term variability, and the proximity of
RX\,J0439 to the theoretical carbon-burning main sequence, raises the
speculative (but spectacular) possibility that RX\,J0439 represents a
completely new type of star.

\acknowledgements{This research was supported by the DLR under grant
50\,OR\,96\,09\,8 and 50\,OR\,99\,03\,6. We thank Falk Herwig for
comments on the evolution of He-burners, Norbert Langer for
interesting discussions, Howard Lanning for technical support with the
HST observations, and the referee, Peter Kahabka, for helpful
comments.}


\begin{thebibliography}{15}
\expandafter\ifx\csname natexlab\endcsname\relax\def\natexlab#1{#1}\fi

\bibitem[\protect\astroncite{{Anderson} et~al.}{1997}]{andersonetal97-1}
{Anderson} S.F., {Margon} B., {Deutsch} E.W., {Downes} R.A., {Allen} R.G.,
  1997, ApJ Lett. 482, L69

\bibitem[\protect\astroncite{{Arons} \& {King}}{1993}]{arons+king93-1}
{Arons} J., {King} I.R., 1993, ApJ Lett. 413, L121

\bibitem[\protect\astroncite{{Bl\"ocker}}{1995}]{bloecker95-1}
{Bl\"ocker} T., 1995, A\&A 299, 755

\bibitem[\protect\astroncite{{Dickey} \& {Lockman}}{1990}]{dickey+lockman90-1}
{Dickey} J.M., {Lockman} F.J., 1990, ARA\&A 28, 215

\bibitem[\protect\astroncite{{G\"ansicke} et~al.}{1998}]{gaensickeetal98-1}
{G\"ansicke} B.T., {van Teeseling} A., {Beuermann} K., {de Martino} D., 1998,
  A\&A 333, 163

\bibitem[\protect\astroncite{{Greiner} et~al.}{1994}]{greineretal94-1}
{Greiner} J., {Hasinger} G., {Thomas} H.C., 1994, A\&A 281, L61

\bibitem[\protect\astroncite{{Hutchings} et~al.}{1995}]{hutchingsetal95-1}
{Hutchings} J.B., {Cowley} A.P., {Schmidtke} P.C., {Crampton} D., 1995, AJ 110,
  2394

\bibitem[\protect\astroncite{{Kippenhahn} \&
  {Weigert}}{1994}]{kippenhahn+weigert90-1}
{Kippenhahn} R., {Weigert} A., 1994, {\em Stellar Structure and Evolution\/}
  (Heidelberg: Springer)

\bibitem[\protect\astroncite{{Rauch}}{1997}]{rauch97-1}
{Rauch} T., 1997, A\&A 320, 237

\bibitem[\protect\astroncite{{van Teeseling}}{1998}]{vanteeseling98-1}
{van Teeseling} A., 1998, in {\em Wild Stars in the Old West: Proceedings of
  the 13th North American Workshop on CVs and Related Objects\/}, {Howell} S.,
  {Kuulkers} E., {Woodward} C. (eds.), pp. 385--394 (ASP Conf. Ser. 137)

\bibitem[\protect\astroncite{{van Teeseling}
  et~al.}{1994}]{vanteeselingetal94-1}
{van Teeseling} A., {Heise} J., {Paerels} F., 1994, A\&A 281, 119

\bibitem[\protect\astroncite{{van Teeseling}
  et~al.}{1996}]{vanteeselingetal96-1}
{van Teeseling} A., {Reinsch} K., {Beuermann} K., 1996, A\&A 307, L49

\bibitem[\protect\astroncite{{van Teeseling}
  et~al.}{1997}]{vanteeselingetal97-1}
{van Teeseling} A., {Reinsch} K., {Hessman} F.V., {Beuermann} K., 1997, A\&A
  323, L41

\bibitem[\protect\astroncite{{Werner}}{1991}]{werner91-1}
{Werner} K., 1991, A\&A 251, 147

\bibitem[\protect\astroncite{{Werner} \& {Wolff}}{1999}]{werner+wolff99-1}
{Werner} K., {Wolff} B., 1999, A\&A 347, L9

\end{thebibliography}
\end{document}